\documentclass[conference]{IEEEtran}
\IEEEoverridecommandlockouts
\usepackage{cite}
\usepackage{amsmath,amssymb,amsfonts}
\usepackage{algorithmic}
\usepackage{graphicx}
\usepackage{textcomp}
\usepackage[table]{xcolor}
\def\BibTeX{{\rm B\kern-.05em{\sc i\kern-.025em b}\kern-.08em
    T\kern-.1667em\lower.7ex\hbox{E}\kern-.125emX}}

\usepackage{multirow}
\usepackage{booktabs}
\usepackage{hyperref}
\usepackage{cleveref}

\crefname{figure}{Fig.}{Figs.}
\crefname{table}{Table}{Tables.}
    
\begin{document}

\title{Satellite Cybersecurity Across Orbital Altitudes: Analyzing Ground-Based Threats to  \\LEO, MEO, and GEO}

\author{\IEEEauthorblockN{Mark Ballard}
\IEEEauthorblockA{The Ohio State University \\
ballard.313@osu.edu}
\and
\IEEEauthorblockN{Guanqun Song}
\IEEEauthorblockA{The Ohio State University \\
song.2107@osu.edu}
\and
\IEEEauthorblockN{Ting Zhu}
\IEEEauthorblockA{The Ohio State University \\
zhu.3445@osu.edu}
}

\maketitle

\begin{abstract}
The rapid proliferation of satellite constellations, particularly in Low Earth Orbit (LEO), has fundamentally altered the global space infrastructure, shifting the risk landscape from purely kinetic collisions to complex cyber-physical threats. While traditional safety frameworks focus on debris mitigation, ground-based adversaries increasingly exploit radio-frequency links, supply chain vulnerabilities, and software update pathways to degrade space assets. This paper presents a comparative analysis of satellite cybersecurity across LEO, Medium Earth Orbit (MEO), and Geostationary Earth Orbit (GEO) regimes. By synthesizing data from 60 publicly documented security incidents with key vulnerability proxies—including Telemetry, Tracking, and Command (TT\&C) anomalies, encryption weaknesses, and environmental stressors—we characterize how orbital altitude dictates attack feasibility and impact. Our evaluation reveals distinct threat profiles: GEO systems are predominantly targeted via high-frequency uplink exposure, whereas LEO constellations face unique risks stemming from limited power budgets, hardware constraints, and susceptibility to thermal and radiation-induced faults. We further bridge the gap between security and sustainability, arguing that unmitigated cyber vulnerabilities accelerate hardware obsolescence and debris accumulation, undermining efforts toward carbon-neutral space operations. The results demonstrate that weak encryption and command path irregularities are the most consistent predictors of adversarial success across all orbits. 
\end{abstract}

% \begin{IEEEkeywords}
% component, formatting, style, styling, insert
% \end{IEEEkeywords}

\section{Introduction}
% (What’s the problem? why it’s an important problem?)
In 2024, the global space environment is more densely populated and commercially critical than at any point in history~\cite{cbo2023constellations}. Over 84\% of active satellites now operate in Low Earth Orbit (LEO), driven largely by the rapid deployment of large constellations for communications, Earth observation, and navigation services~\cite{cbo2023constellations}. This shift has fundamentally changed both the technical and risk landscapes in orbit. While traditional space safety research has focused on kinetic threats such as collisions and debris, the parallel rise of cyber-physical threats to satellites has created a new category of systemic risk: space-based infrastructure that can be degraded or commandeered remotely from Earth.

Unlike debris impacts, many satellite failures and degradations can be caused, amplified, or coordinated from the ground through cyber and electromagnetic means. Ground-based adversaries can target satellites via their radio-frequency (RF) links, ground segment interfaces, supply-chain vulnerabilities, and software update pathways~\cite{nist2022cybersecurity}. Tactics such as jamming, spoofing, and protocol exploitation do not require access to space; they exploit the fact that satellites must remain communicative and responsive to ground commands to be useful~\cite{smith2024jamming}. 

Moreover, emerging communication modalities introduce new attack vectors; vulnerabilities in optical communication links have been identified~\cite{290983}, and adversarial attacks on machine-learning-based low-power protocols can disrupt critical telemetry links~\cite{song2022mlbasedsecurelowpowercommunication}. As a result, adversaries can attack space systems without ever leaving Earth’s surface.

However, not all satellites are equally exposed. Cybersecurity challenges vary significantly across orbital altitudes. LEO systems tend to be numerous, commercially driven, and rapidly iterated, often relying on frequent contact with ground stations and user terminals. 
These dense constellations also face unique physical constraints; optimizing energy efficiency for communication protocols like LoRaWAN in LEO is critical for operational longevity~\cite{shergill2024energyefficientlorawanleo}, while managing thermal loads in on-board computing units is essential to prevent hardware failure~\cite{yuan2024heatsatellitesmeatgpus}.
Medium Earth Orbit (MEO) satellites, particularly those in navigation constellations, have unique timing and integrity requirements that make them attractive for spoofing attacks~\cite{slater2024megaconstellations}. Geostationary Earth Orbit (GEO) satellites, while few in number and physically distant, provide globally critical services; their high altitude influences the physics, feasibility, and detectability of attacks~\cite{slater2024megaconstellations}.

Beyond immediate operational risks, cyber-induced failures have broader sustainability implications. Hardware obsolescence caused by unpatchable vulnerabilities or thermal runaway contributes to the growing debris problem, contradicting efforts towards carbon-neutral I/O device lifecycles~\cite{yu2024achievingcarbonneutralityio, cheng2024technologicalprogressobsolescenceanalyzing}.

Understanding how distance and orbital dynamics impact both attack feasibility and effectiveness is crucial for building resilient satellite ecosystems. This paper analyzes satellite cybersecurity across three primary orbital regimes—LEO, MEO, and GEO—with a particular emphasis on ground-based RF and cyber threats. We focus on how orbit altitude affects the feasibility of cyber attacks. We then synthesize these differences into altitude-aware defensive strategies, arguing that space cybersecurity cannot be treated as a uniform problem. Instead, it demands tailored protections aligned with each orbit’s communication patterns, mission profiles, and physical constraints.

\section{Related Work}
Early space security literature concentrated on physical threats, especially collisions and debris generation, exemplified by the Kessler Syndrome concept and subsequent modeling of debris cascades~\cite{kessler1978collision}. Over time, research expanded to include intentional kinetic threats, such as anti-satellite (ASAT) weapons, but cyber and RF-based threats to satellites have lagged behind in both regulation and academic modeling~\cite{csis2023antisatellite}.

Recent work has started to bridge this gap. Several strands of research are relevant to satellite cybersecurity across orbital altitudes. First, studies of GNSS (Global Navigation Satellite System) vulnerabilities have highlighted the susceptibility of satellite navigation signals to jamming and spoofing, particularly at the user-equipment level~\cite{humphreys2008spoofing}. Experimental demonstrations have shown that relatively low-cost ground-based transmitters can overpower legitimate GNSS signals, misleading receivers about their position and time~\cite{humphreys2008spoofing}. 

Furthermore, as satellite networks increasingly integrate with terrestrial infrastructure, vulnerabilities in 5G networks, such as location tracking risks, become relevant to hybrid satellite-ground architectures~\cite{ali2023security5gnetworks}.
These works focus primarily on MEO-based systems (e.g., GPS, Galileo) but often treat the space segment as a given rather than as a cyber target in its own right.

Second, research on satellite communications security has documented cases of unintentional and intentional interference affecting GEO communications satellites. Regulatory bodies and monitoring organizations have reported instances of signal hijacking, deliberate uplink interference, and unauthorized use of transponders~\cite{wired2009fleetcom}. 

To mitigate these risks in future networks, researchers are exploring advanced frameworks, such as optimizing global quantum communication via satellite constellations~\cite{gao2024optimizingglobalquantumcommunication} and employing heterogeneous computing systems to enhance on-board data processing security~\cite{khatri2022heterogeneouscomputingsystems}.
Much of this literature is descriptive, cataloging incidents and regulatory responses, but provides limited systematic analysis of how orbit altitude constrains these attacks.

\begin{figure*}[t]
    \centering
    \includegraphics[width=0.8\textwidth]{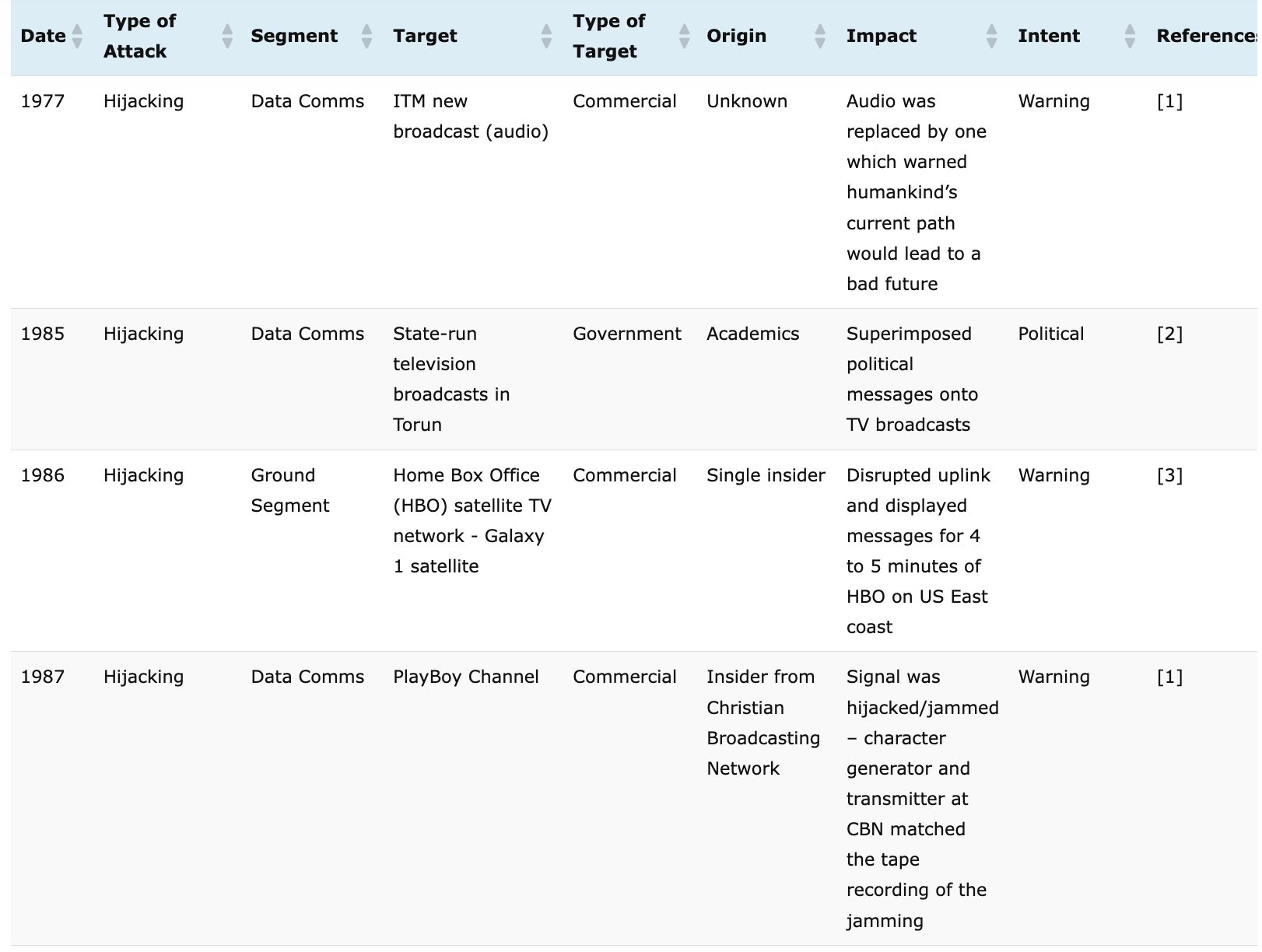}
    \caption{
        Space \& Cybersecurity Info Space Attacks Open-Database
    }
    \label{fig:1}
   
\end{figure*}
Third, emerging cybersecurity work on LEO mega-constellations has raised concerns about systemic risk~\cite{ucsd_satcom_sysnet}. Analysts have argued that large, software-defined constellations introduce a massive attack surface: thousands of nearly identical satellites, frequent over-the-air updates, and extensive use of ground and user-side software~\cite{ucsd_satcom_sysnet}.

The software stack itself presents numerous vulnerabilities. Security challenges in microservices architectures~\cite{gopal2022securityprivacychallengesmicroservices} and OS-level virtualization~\cite{ketha2025analysissecurityoslevelvirtualization} are increasingly critical as satellites adopt cloud-native software designs. Moreover, the integrity of the ground segment relies on secure internal mechanics, such as those found in Windows security environments~\cite{kulshrestha2023innerworkingswindowssecurity} and robust virtual file systems~\cite{sun2023designimplementationconsiderationsvirtual}. To manage the immense data flow from these constellations securely, efficient multiprocessing strategies for data classification~\cite{dixit2023dataclassificationmultiprocessing} and map-reduce tasks~\cite{qiu2023mapreducemultiprocessinglargedata} are essential. Some researchers have also proposed blockchain scalability solutions to ensure verifiable command logs across decentralized satellite networks~\cite{li2022minisculesurveyblockchainscalability}.
A single exploited vulnerability in the control software or networking stack could cascade across dozens or hundreds of spacecraft~\cite{smith2024jamming}.

There is, therefore, a gap between specific, incident-focused research (e.g., GNSS spoofing experiments, GEO interference cases, and LEO constellation risk analyses) and broader modeling that differentiates threats by orbital altitude. This paper addresses that gap by explicitly analyzing ground-based cyber and RF threats as a function of orbit.

\section{Design}

Our study was designed to compare ground-based cyber and RF threats across LEO, MEO, and GEO by combining a curated incident database with orbit-level vulnerability proxies derived from historical anomaly studies. The objective of this design was to characterize how often satellites in different orbits experience operational conditions that could make them more susceptible to jamming, hijacking, and other ground-initiated attacks, and to clarify how those conditions differ across orbital regimes. 

\subsection{Data Sources}

The primary incident source was the Space \& Cybersecurity Info space attacks open database, which we used as the basis for this study. The dataset compiles 60 publicly documented incidents involving interference with space systems between 1977 and 2019, drawn from open-source reporting and community curation. Each record includes the date, type of attack (for example, jamming, broadcast hijacking, computer network exploitation or eavesdropping), the target satellite or ground system, the reported origin when available, the observed impact (such as signal disruption, loss of service, or data compromise), and references. For our orbit comparison, we additionally coded each incident by orbital regime (LEO, MEO, GEO) and by a category label intended to capture the most salient enabling condition described in reporting, such as high uplink activity, anomalous TT\&C behavior, weak encryption, or power-related anomalies. Representative examples include the APSTAR 1A broadcast hijacking in 1997 (GEO, unauthorized override of television content), GPS jamming reported during the Iraq conflict in 2003 (MEO, localized navigation disruption), and the 1998 ROSAT takeover allegation (LEO, suspected unauthorized command access)~\cite{space_attacks_db}. 

\subsection{Vulnerability Proxies}

To characterize background vulnerability by orbit, we also drew on historic anomaly and reliability studies that report how frequently spacecraft in different regimes exhibit specific operational and environmental stressors. Because detailed, orbit-resolved cyber telemetry is rarely public, these studies provide a practical way to approximate the conditions that can make ground-based interference more feasible or more damaging. We selected proxies that map to plausible attack pathways and that also appear repeatedly in incident narratives. These include high uplink frequency (including frequent commanding and high-rate uplink operations), TT\&C anomaly as an indicator of control-path fragility, instances where encryption weak (weak or absent encryption can enable interception, replay, or unauthorized commanding), radiation-driven susceptibility such as single-event upsets (SEU) rate high that can degrade fault tolerance, and power anomalies that can convert transient disruptions into sustained loss of service. The proxies are used comparatively across LEO, MEO, and GEO, with the intent of highlighting how orbital regime changes both the technical attack surface and the operational context in which attacks are attempted and detected. 

\subsection{Analytical Approach}

Our analysis combines the incident database with the orbit-level proxy indicators in order to compare both observed attack patterns and the background conditions that can enable them. First, we summarize incidents by orbit and attack type to identify which classes of events appear most frequently within each regime. Second, we use the proxy categories as an interpretive layer to connect incidents to recurring enabling conditions, emphasizing consistency across multiple events.

\begin{figure}[t]
  \includegraphics[width=1.0\columnwidth]{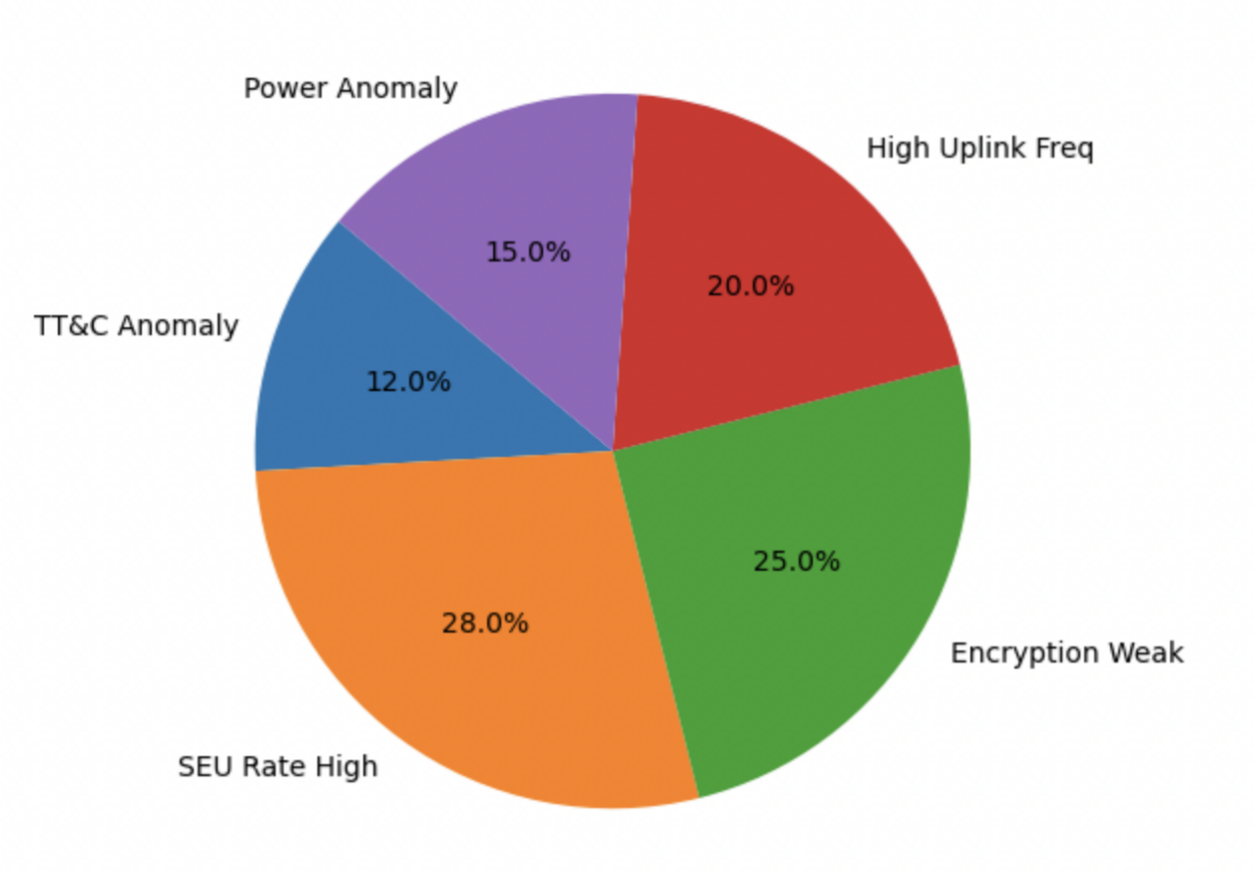}
  \caption{Five Common Selected CIA-aligned Satellite Vulnerabilities}
  \label{fig:2}
\end{figure}

\begin{figure}[t]
  \includegraphics[width=1.0\columnwidth]{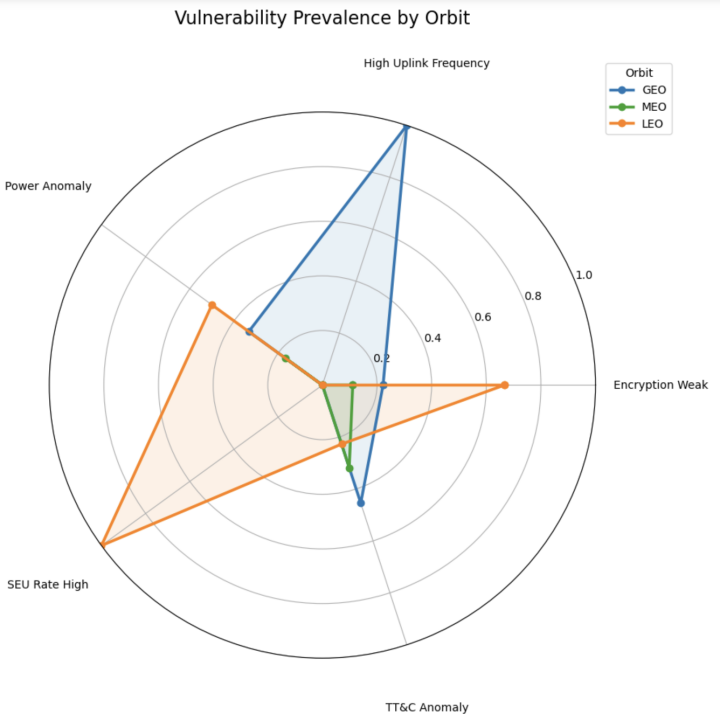}
  \caption{Prevalence of each vulnerability across LEO, MEO, and GEO from the open-source database 60 publicly documented incidents}
  \label{fig:3}
\end{figure}

\section{Evaluation}
Figure \ref{fig:1} summarizes this incident record by orbit and shows that reported events cluster most heavily in GEO with fewer incidents attributed to MEO and LEO. This skew toward GEO is consistent with the concentration of high-value communications broadcast and military missions in geostationary orbit. It also reflects the operational reality that GEO systems provide persistent coverage and stable link geometries that can make interference and command path abuse attractive to adversaries.

Building on this incident baseline the evaluation focuses on vulnerability proxies. These are indicators of key cyber-physical weak points in satellites rather than random hardware failures. In the dataset these proxies are captured by category labels assigned to each incident. Figure 2 shows an approximate distribution of these categories derived by synthesizing trends from the literature. The resulting estimated shares are SEU Rate High as the most common proxy at 28\%, followed by Encryption Weak at 25\%, High Uplink Frequency at 20\%, Power Anomaly at 15\%, and TT\&C Anomaly at 12\%. These figures were obtained by combining and normalizing prominent quantitative patterns. Examples include the dominance of radiation-induced upsets in solar events~\cite{nasa2005fragmentations}, the high prevalence of unencrypted traffic in recent scans~\cite{ccs2025dontlookup}, inferred risks from uplink monitoring priorities~\cite{breakingdefense2024spaceforce}, historical correlations with power subsystem degradation~\cite{german_lib_report}, and scaled contributions from guidance, navigation, and control anomalies~\cite{nasa2003debris}.

These categories map to practical pathways seen in historical incidents and technical analyses. TT\&C Anomaly reflects irregular behavior on the command and telemetry plane, which mission assurance reporting treats as operationally significant, because it can indicate unexpected state changes or off-nominal commanding and requires careful investigation and mitigation in spacecraft operations~\cite{nasa2003debris}. SEU Rate High reflects the well-established space environment risk that radiation can upset electronics and memory, producing transient faults that may cascade into larger anomalies and complicate incident interpretation when environmental effects overlap with suspicious behavior~\cite{nasa2005fragmentations}. Encryption Weak denotes insufficient confidentiality and authentication on satellite links, a recurring theme in SATCOM security research that emphasizes how weak or missing cryptography can enable interception, spoofing, and unauthorized control attempts~\cite{ccs2025dontlookup}. High Uplink Frequency captures the practical exposure created by frequent uplink interactions and a dense contested RF environment, where scale and monitoring challenges increase the number of opportunities for interference or misuse of access paths~\cite{breakingdefense2024spaceforce}. Power Anomaly reflects instability or degradation in the electrical power subsystem, a known driver of safe mode transitions, subsystem resets, and reduced operational margins that can amplify the impact of other disturbances and create denial or degradation outcomes~\cite{german_lib_report}.

\begin{figure}[t]
  \includegraphics[width=1.0\columnwidth]{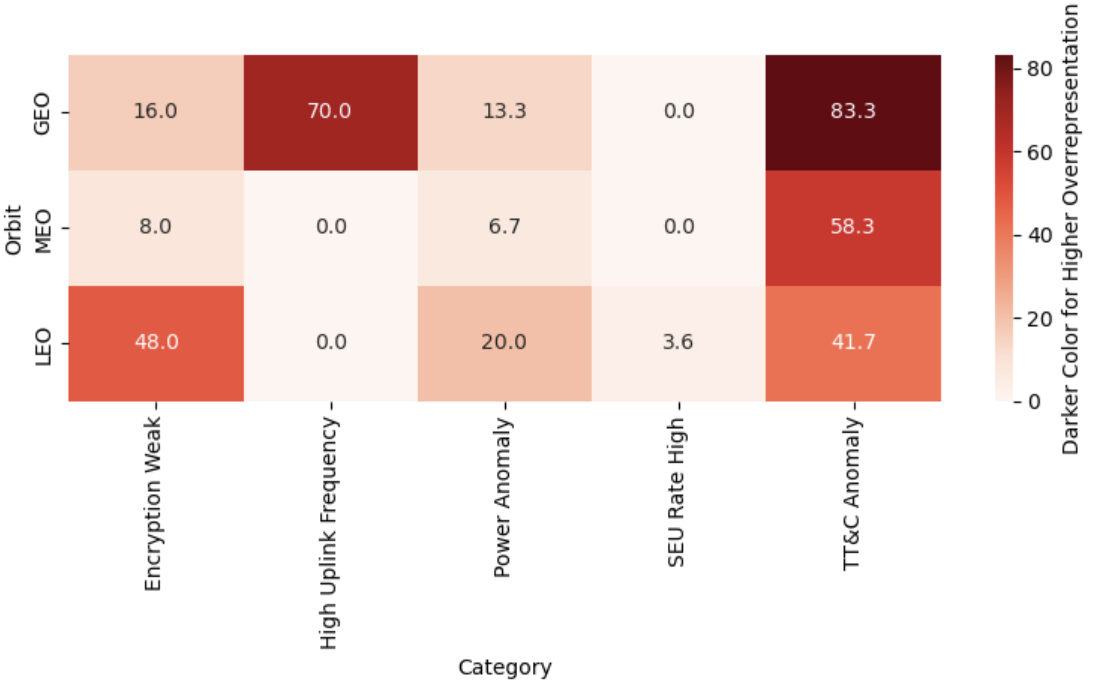}
  \caption{Evaluation of Vulnerabilities to Attacks that are Overrepresented by Orbit}
  \label{fig:4}
\end{figure}

Because overall frequency can mask orbit-specific exposure, the analysis also compares category patterns by orbit. Figure 3 shows vulnerability prevalence by orbit from the open-source database.  The analysis indicates that GEO is most strongly associated with High Uplink Frequency and a meaningful presence of TT\&C Anomaly, while LEO shows stronger alignment with SEU Rate High and Power Anomaly along with a measurable presence of Encryption Weak. MEO shows more emphasis on control conditions such as TT\&C Anomaly. This orbit differentiation supports the central argument that altitude shapes both how systems are operated and how adversaries can interact with them. 

To move beyond counts, the study evaluates conditional attack association by asking which categories are overrepresented with attack-labeled events within each orbit. Figure 4 presents this as a heatmap of overrepresentation of attack likelihood by vulnerability and orbit, and shows that  TT\&C Anomaly and Encryption Weak are the most consistently overrepresented attack-associated conditions across GEO, MEO, and LEO. In other words, when weak or absent encryption is present, the probability that the observed event reflects adversarial action is elevated across all three orbits. TT\&C Anomalies also show strong association, particularly in GEO and MEO where command links are heavily utilized and frequently contested. Conversely, SEU Rate High is the least predictive when considered in isolation.

Overall, the incident distribution in Figure 1, the category frequency profile in Figure 2, the vulnerability prevalence from the open-source database 60 publicly documented incidents in Figure 3, and the conditional association results in Figure 4 together yield a structured, data-driven view of cyber-physical risk in space systems. The results support orbit-tailored mitigation priorities with consistent emphasis on modern cryptography and on monitoring and hardening of the command and control plane.

\section{ISSUES WE ENCOUNTERED}

Several issues were encountered in the course of this study that affected both data selection and the design of our analysis. One major challenge was the absence of a dedicated TLE dataset that explicitly labels satellites by cyber threat exposure. Publicly available TLE catalogs provide accurate orbital elements, but they do not contain fields indicating whether a satellite has ever been jammed, spoofed, or hacked. Because of this gap, it was not possible to perform a direct, large-scale statistical analysis linking individual TLE records to specific ground-based attack events. A related issue was that the incident data itself is limited and uneven. The open-source Space \& Cybersecurity Info database contains only around 60 events spread over more than four decades, and these are restricted to incidents that were publicly reported, discussed in open literature, or later declassified.

We also found that there is no single, unified dataset that breaks down subsystem anomalies (for example, TT\&C, power, etc.). Instead, relevant information is scattered across separate studies and catalogs, each covering different time windows, satellite classes, and reporting standards. As a result, our vulnerability proxy values had to be derived by combining and normalizing statistics from multiple sources.

Finally, because the number of known incidents is small and the proxy metrics are aggregated, our results are necessarily coarse. They are well-suited to orbit-level comparisons (LEO vs. MEO vs. GEO) and broad trends but are not suitable for predicting the risk profile of any individual satellite. Despite these issues, the combination of incident records and proxy metrics still provides useful insight into how orbital regime, subsystem behavior, and security posture interact to shape the landscape of ground-based threats.

\section{Conclusion}

This work set out to characterize how cyberattacks against space systems present across orbital regimes and how those events relate to recurring technical weaknesses that shape confidentiality integrity and availability outcomes. Using publicly available incident reporting as an empirical baseline and layering a vulnerability centric interpretation on top the study shows that cyber risk in space is not evenly distributed and it is not driven by a single subsystem in isolation. Instead risk emerges from the interaction between orbit specific operations adversary opportunity and the security posture of command and mission links.

A central finding is that some conditions are repeatedly associated with attack realization. Weak encryption stands out as the most consequential enabler because it reduces the cost of interception spoofing and unauthorized command while increasing the probability that anomalous behavior reflects malicious action rather than random error. Closely related are irregularities in the command and control plane where TT\&C anomalies align strongly with attack labeled events and should be treated as potential indicators of compromise rather than as routine telemetry noise. Together these findings reinforce that cryptographic modernization and command path assurance are baseline requirements for resilient space architectures and they must be implemented end to end across payload and control channels. 

Looking ahead, the findings suggest several concrete directions for improvement. At the technical level, space operators should prioritize cryptographic modernization, including deprecating unencrypted or weakly encrypted links, implementing robust key management, and ensuring that both payload and TT\&C channels are covered by strong, updatable cryptographic schemes. In parallel, intelligent anomaly detection systems, including machine learning models trained on TT\&C and telemetry data, could help distinguish between benign anomalies (e.g., SEUs, hardware aging) and signatures consistent with known attack patterns. Such systems would be particularly valuable in all orbit levels, where TT\&C anomalies and encryption weaknesses have the highest observed attack correlation.

\bibliography{reference,zhu}
\bibliographystyle{ieeetr}

\end{document}